\documentclass[a4paper,twocolumn,accepted=2024-07-26]{quantumarticle}
\pdfoutput=1

\usepackage{graphicx}% Include figure files
\usepackage[numbers]{natbib}
\usepackage{dcolumn}% \documentclass[aps,prl,reprint,superscriptaddress,nobibnotes]{revtex4-2}
\usepackage{blindtext}
\usepackage{lineno}
\usepackage{amsmath}
\usepackage{epstopdf}
\usepackage{float}
\usepackage{txfonts}
\usepackage[utf8]{inputenc}
\usepackage[T1]{fontenc}

\usepackage{esint}
\usepackage{braket}
\usepackage{hyperref}

\usepackage[separate-uncertainty=true]{siunitx}

\usepackage[dvipsnames]{xcolor}

\colorlet{mylinkcolor}{RoyalPurple}
\colorlet{mycitecolor}{RoyalPurple}
\colorlet{myurlcolor}{RoyalPurple}
\usepackage{hyperref}
\hypersetup{
	linkcolor  = mylinkcolor,
	citecolor  = mycitecolor,
	urlcolor   = myurlcolor,
	colorlinks = true,
	breaklinks = true
}

\DeclareSIUnit{\rad}{rad}
\DeclareSIUnit{\px}{px}

\newcommand{\subfig}[1]{(#1)}
\newcommand{\iu}{{i\mkern1mu}}

\begin{document}
\title{Long-lived collective Rydberg excitations in atomic gas achieved via ac-Stark lattice modulation}
\author{Stanisław Kurzyna}
\thanks{Equal contributions}
\affiliation{Centre for Quantum Optical Technologies, Centre of New Technologies, University of Warsaw, Banacha 2c, 02-097 Warsaw, Poland}
\affiliation{Faculty of Physics, University of Warsaw, Pasteura 5, 02-093 Warsaw, Poland}
\author{Bartosz Niewelt}
\orcid{0009-0005-8710-8077}
\thanks{Equal contributions}
\affiliation{Centre for Quantum Optical Technologies, Centre of New Technologies, University of Warsaw, Banacha 2c, 02-097 Warsaw, Poland}
\affiliation{Faculty of Physics, University of Warsaw, Pasteura 5, 02-093 Warsaw, Poland}
\author{Mateusz Mazelanik}
\email{m.mazelanik@cent.uw.edu.pl}
\affiliation{Centre for Quantum Optical Technologies, Centre of New Technologies, University of Warsaw, Banacha 2c, 02-097 Warsaw, Poland}
\author{Wojciech Wasilewski}
\affiliation{Centre for Quantum Optical Technologies, Centre of New Technologies, University of Warsaw, Banacha 2c, 02-097 Warsaw, Poland}
\affiliation{Faculty of Physics, University of Warsaw, Pasteura 5, 02-093 Warsaw, Poland}
\author{Michał Parniak}
\orcid{0000-0002-6849-4671}
\email{m.parniak@cent.uw.edu.pl}
\affiliation{Centre for Quantum Optical Technologies, Centre of New Technologies, University of Warsaw, Banacha 2c, 02-097 Warsaw, Poland}
\affiliation{Faculty of Physics, University of Warsaw, Pasteura 5, 02-093 Warsaw, Poland}

\begin{abstract}
Collective Rydberg excitations provide promising applications ranging from quantum information processing, and quantum computing to ultra-sensitive electrometry. However, their short lifetime is an immense obstacle in real-life scenarios. The state-of-the-art methods of prolonging the lifetime were mainly implemented for ground-state quantum memories and would require a redesign to effectively work on different atomic transitions. We propose a protocol for extending the Rydberg excitation lifetime, which in principle can freeze the spin-wave and completely cancel the effects of thermal dephasing. The protocol employs off-resonant ac-Stark lattice modulation of spin waves by interfering two laser beams on the atomic medium. Our implementation showed that the excitation lifetime can be extended by an order of magnitude, paving the way towards more complex protocols for collective Rydberg excitations.
 \end{abstract}
\maketitle
%%%%%%%%%%%%%%%%%%%%%%%%%%  body  %%%%%%%%%%%%%%%%%%%%%%%%%%

\section{Introduction}
\begin{figure}[t]
    \centering
    \includegraphics[width=0.95\columnwidth]{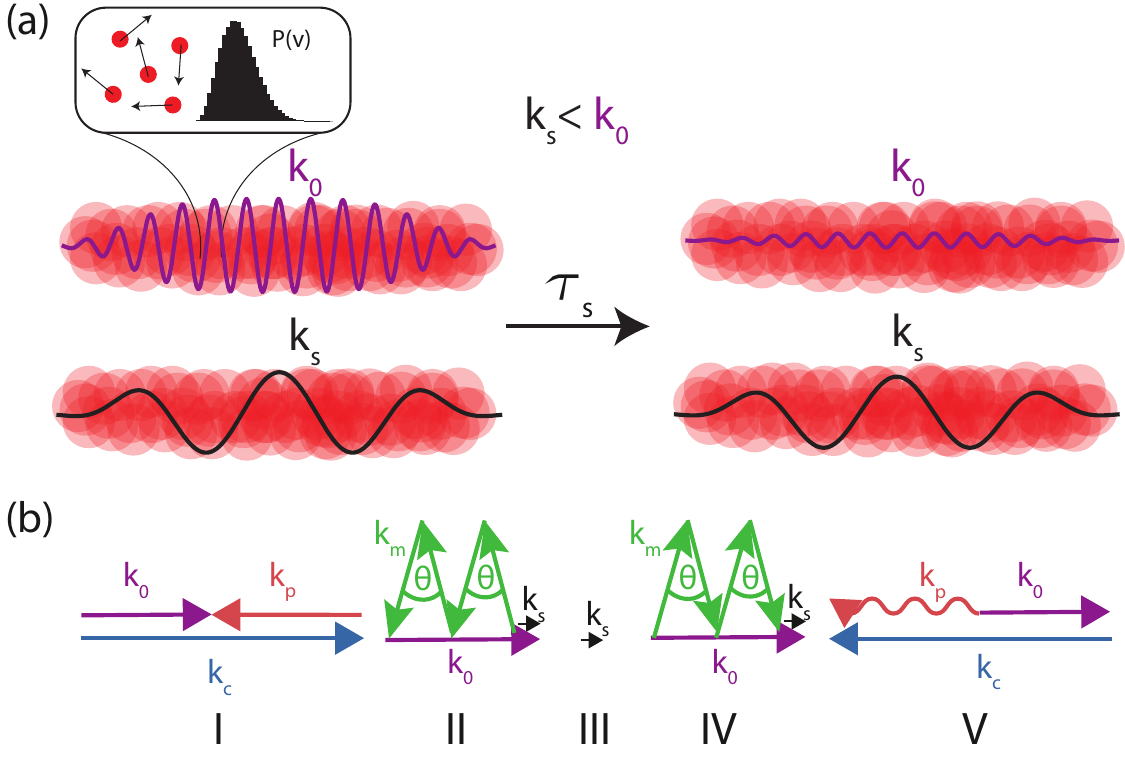}
    \caption{
    \subfig{a} Ideational representation of the blurring of the spin waves with different wavevectors. Due to non zero temperature of the ensemble, atoms have a residual velocity following the Maxwell-Boltzman distribution. Spin waves with larger wavevector are more affected by motional dephasing, which corresponds to shorter thermal lifetime. 
    \subfig{b} Geometrical representation of the stages of the extended lifetime protocol. (I) - Probe ($\text{k}_\text{p}$) and coupling ($\text{k}_\text{c}$) laser fields excite atomic spin wave ($\text{k}_\text{0}$), (II) - Stored spin wave is modulated with two crossed beams ($\text{k}_\text{m}$), (III) - Spin wave ($\text{k}_s$) with longer lifetime is stored, (IV) - Modulation is repeated to revert the spin wave to its original state, (V) - Read-out with coupling laser field.
    }
    \label{fig:idea}
\end{figure}

Atomic ensembles of highly excited Rydberg states are subjects of vastly increasing interest due to their key role in quantum technologies.     
Their application varies from ultra sensitive sensing \cite{borowka_continuous_2023,Borowka:22,Dietsche2019,Sedlacek2012} through quantum computing \cite{Cohen2021,Ryabtsev2005,Saffman2016}, quantum simulation \cite{Scholl2021,Weimer2010,nguyen_towards_2018,Lukin2001} and quantum information processing \cite{ParedesBarato2014,Saffman2010,Zhu2022} to the single photon generation \cite{Heller2022,Ripka2018,Yang2022} and generation of high-fidelity entanglement \cite{Li2013,Saffman2002}, by using Floquet frequency modulation \cite{Zhao2023} even outside the blockade radius. 
Recently, storage of a photon as a collective excitation of the ground state and Rydberg-level state atoms i.e. spin wave, has been ardently explored by examining Rydberg polaritons \cite{Jia2018,Ripka2016,maxwell_storage_2013,Drori2023} and employing them in quantum memories \cite{Distante2016,dudin_strongly_2012,lowinski_strongly_2024,spong_collectively_2021}. 
Most of the aforementioned experiments have relied on coherence between ground and Rydberg states. 
However, to access the Rydberg state optical fields of distinctly different wavelengths are employed which leads to a large wavevector of the produced spin wave and thus quick thermal decoherence as depicted in Fig. \ref{fig:idea}. This leads to very short memory lifetimes, i.e. of an order of a few microseconds, and prevents implementations of in-memory operations. 
Prolonging the memory lifetime seems advantageous and may lead to a more robust application of Rydberg states. 
The solutions for mitigating the motional dephasing may be based on zeroing the spin-wave momentum \cite{jiang_dynamical_2016} or employing an optical lattice \cite{Dudin2013,lampen_long-lived_2018,Zhao2009}. Still, those methods were implemented for ground state memory or by implementing a magic wavelength lattice trap for ground-state-Rydberg coherence.

Another approach can be implemented with the use of continuous dressing fields \cite{PhysRevX.11.011008} or by storing the information in the collective excitation of the ground state and transferring the coherence to Rydberg level \cite{li_quantum_2016} only for a short period. Despite providing a feasible way of conserving properties of Rydberg polaritons for a longer time it requires auxiliary intermediate state.

Recent advances in spin-wave engineering and manipulation \cite{Li2011,Xu2013}, especially spatial spin-wave modulation technique \cite{Niewelt2023,Mazelanik2022} allow for significant modifications of longitudinal or perpendicular \cite{parniak_quantum_2019,Jastrzebski2024} wavevector. Remarkably, as depicted in Fig. \ref{fig:idea}\subfig{b} by applying a proper phase pattern, the spin-wave wavevector can be shrunk so that the thermal dephasing plays a negligible role in the memory efficiency, vastly increasing storage time.

In this article, we present a novel and robust protocol for extending the lifetime of Rydberg quantum memory by engineering the momentum of the ground-Rydberg state coherence via spatial lattice modulation. By using two interfering beams we were able to freeze the thermal motional decoherence and show that in principle it allows for an arbitrarily long lifetime, limited only by the radiative lifetime of the excited state $1/(2\pi\Gamma) =  \SI{24.1}{\us}$ \cite{Branden2009}. 
In contrast to previously presented methods employed in the $\Lambda$-system level configuration \cite{jiang_dynamical_2016,li_quantum_2016}, our protocol is compatible with ladder type memories \cite{kaczmarek_high-speed_2018,Davidson2023,Finkelstein2018} and especially Rydberg quantum memories and as it requires modulating only the ground state of the induced atomic coherence. 
Long-lived Rydberg quantum memory may be employed in many quantum information processing protocols by employing microwave modulation \cite{Fan:23,Liu2022} or spatial spin-wave modulation technique. In particular, long storage of qubits collectively encoded in an atomic ensemble may provide significant improvement in quantum computing and quantum networking \cite{Kimble2008,Leent2020}.

\section{Theory}
Ground-Rydberg coherence coupled to optical fields will typically be produced with significant momentum i.e. wavevector. Therefore the phase of coherence $\rho_{gr}(z) \propto \exp(i k_0 z)$ varies quickly. The residual motion of atoms in a cold ensemble in typical experiments $T\approx \SI{100}{\micro K}$  at $v_\text{t}=\sqrt{k_B T / m}$ is fast enough to blur such spin waves in a matter of microseconds, exactly $\tau=(k_0 v_\text{t})^{-1}$. Ideational scheme of the effect is presented in Fig.~\ref{fig:idea}\subfig{a}.
Extending the thermal lifetime requires an operation that shortens the spin-wave wavevector $k_0$. Ideally, we would cancel the phase $\exp(i k_0 z)$ of the spin wave via a sub-microsecond linear phase modulation. However, such a large light shift is infeasible. Applying an intensity ramp along $z$ would require enormous laser intensities to counteract tens of thousands of radians of $k_0 z$. A Fresnel i.e. periodic ramp pattern could reduce the intensity needed for successful modulation. However, the ramp period $2\pi/k_0 \approx \SI{1.2}{\um}$ is so short that it couldn't be produced optically with high fidelity.

Instead, we opted for a solution that uses just two interfering laser beams that form an optical lattice and utilizes only half of the base frequency of the periodic ramp pattern -- a sinusoidal phase modulation $e^{i A \sin \left(q z\right) t}$. Using only half of the base frequency results in the stored coherence transfer to a shorter wavevector state in the second order of spin-wave diffraction. A simplified geometric sketch of coherence wavevector at each step of the protocol is presented on Fig.~\ref{fig:idea}\subfig{b}. The initial wavevector of the coherence is generated in step (I). Later the modulation is applied in the second order (II), the coherence is then transferred to a state with much longer lifetime (III). The modulation has to be reapplied (IV) so the coherence can be read out in the last step (V).

 \begin{figure}[t]
    \centering
    \includegraphics[width=0.95\columnwidth]{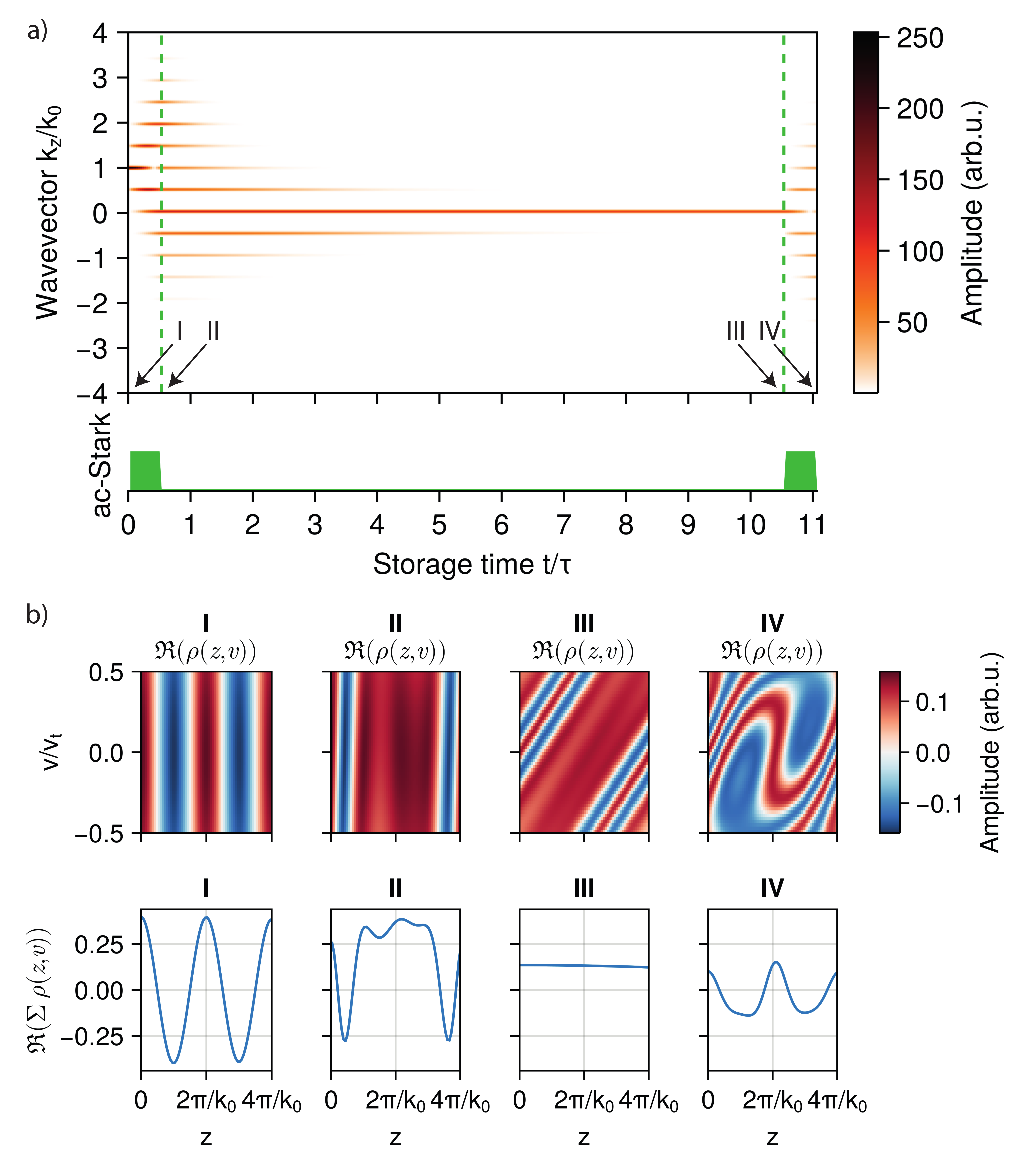}
    \caption{
    \subfig{a} The simulated time dependence of Fourier components of the spin-wave coherence with $10\tau$ delay between modulation, in case of sinusoidal modulation during moments: I to II and III to IV. The green bars below the main plot correspond to moments when the modulation was active. Note that the modulation generates residual components with non-zero wavevector that decay after some time 
    \subfig{b} In the first row, real part of phase-space density in moments I, II, III, and IV. In second row real part of the phase-space density averaged over velocity classes. The stored oscillation in plot (I) is modulated with sinusoidal phase pattern (II). After time, the coherence evolves differently for each of the velocity classes (III). At the end, the oscillation reappears following the second modulation (IV), long after it would decay without the modulation.
    }
    \label{fig:simvw}
\end{figure}

Under this operation, with modulation along the Z axis, the wavevector representation of spin wave $\tilde{\rho}(\boldsymbol{k})$ is transformed according to:
\begin{equation}
\begin{split}
    \tilde{\rho}_\text{mod} (\boldsymbol{k})  &= \int  \rho(\boldsymbol{r})e^{i \boldsymbol{k}\cdot\boldsymbol{r}} e^{i A \sin \left(q z\right) t} \mathrm{d}\boldsymbol{r} \\
    &= \int \sum_{n=-\infty}^{\infty}  J_{n}(A t) \rho(\boldsymbol{r}) e^{\iu  \left( \boldsymbol{k}\cdot\boldsymbol{r} - n q z\right)} \mathrm{d} \boldsymbol{r} \\
    &= \sum_{n=-\infty}^{\infty} J_{n}(A t) \, \tilde{\rho}\left(\boldsymbol{k} - n q \hat{\boldsymbol{k_z}}\right),
\label{eq:bessel}
\end{split}
\end{equation}

where $J_{n}$ is $n$-th order Bessel function of the first kind. Let us assume a typical wavevector distribution of the coherence defined by the spatial distribution of the atoms. Such distribution $\tilde{\rho}$ is centered around 
 $\boldsymbol{k_0} = k_0 \boldsymbol{\hat{z}}$ oriented in the Z direction. The modulation splits the distribution into diffraction orders with amplitudes given by $J_{n}(A t)$. A zero-momentum order centered around $\boldsymbol{k}=0$ appears if the modulating wavenumber $q$ fulfills the equation $nq = k_0$ for some natural number $n$. Such precise alignment of the phase modulation could freeze the spin wave and halt the effects of thermal decoherence. In our case, we utilize the second order, thus $n=2$ and $q = k_0/2$.
 This limits the overall retrieval efficiency to $\eta_{\text{max}} =  \text{max}[J_{2}(t)]^4 \approx \qty{5.6}{\percent}$, as two modulations are needed, one after write-in and one just before read-out.
 
\section{Simulation}
To better understand the protocol limitations, we simulated the coherence behavior by considering a 1D phase-space evolution of the atomic medium. 
For simplicity, we considered the evolution of the stored plane spin wave $e^{\iu k_0 z}$ and assumed a smooth spatial atomic density following a Gaussian distribution with a standard deviation equal to $15\pi / k_0$. The distribution in velocity is a Gaussian with standard deviation equal $v_\text{t}$. As a time unit, we assumed the lifetime of the non-modulated spin wave $\tau = (k_0 v_\text{t})^{-1}$. In the simulation, the $z$-axis was divided into $2^{11}$ points, ranging from $-4L$ to $4L$, the velocities were ranging from $-4v_\text{t}$ to $4v_\text{t}$ and were divided into 400 points.
The phase-space distribution at every point in time experiences evolution corresponding to a shift in position proportional to the current value of velocity and time difference between each step. In the beginning, we store the spin wave and modulate the ensemble by imprinting a position-dependent phase $e^{i A \sin \left(q z\right)}$ for time $\tau_\text{acS} = 0.54 \: \tau$. The modulation parameter $q=0.485 \: k_0$ is set to reflect the imperfections of the experimental setup's alignment. Then after some delay, we re-apply the modulation for the same duration and average the final distribution over all classes of velocities and over all positions with counter-rotating term $e^{-\iu k_0 z}$. At each time step the distribution is multiplied by an exponential term to reflect the spontaneous decay rate $\Gamma$ of the selected Rydberg state \cite{Branden2009}. The expected read-out intensity is squared absolute value of the averaged distribution.

The map of Fourier components of the phase-space density averaged over velocities is presented in Fig.~\ref{fig:simvw}\subfig{a}. It allows us to take a peek into the spin-wave evolution under the proposed protocol and observe the behavior of different components of the coherence. The stored wave with $k_0$ (I) is modulated and transferred to a state with a longer lifetime (II). The time of the modulation is optimized so that the amplitude of the component with the lowest wavevector is maximal. However, the modulation generates multiple of higher order (in wavevector) components of the coherence, which decay after some time ($\approx5\tau$). Later, the modulation is reapplied (III) and the wave with $k_0$ is partially revived (IV). The images in the first row of Fig.~\ref{fig:simvw}\subfig{b} show the real part of the phase-space distribution in different moments of the protocol (denoted with Roman numerals). The second row plots the real parts of the phase-space distribution averaged over velocities in the corresponding moments in time. The oscillations present at the beginning of the protocol (I) reappear at the end (IV) long after they would decay without the modulation. This shows that the spin wave was temporarily transferred to a lower-wavevector state, effectively extending the lifetime of the excitation.

\section{Experimental setup}
\begin{figure}[t]
\centering
\includegraphics[width = 1\columnwidth]{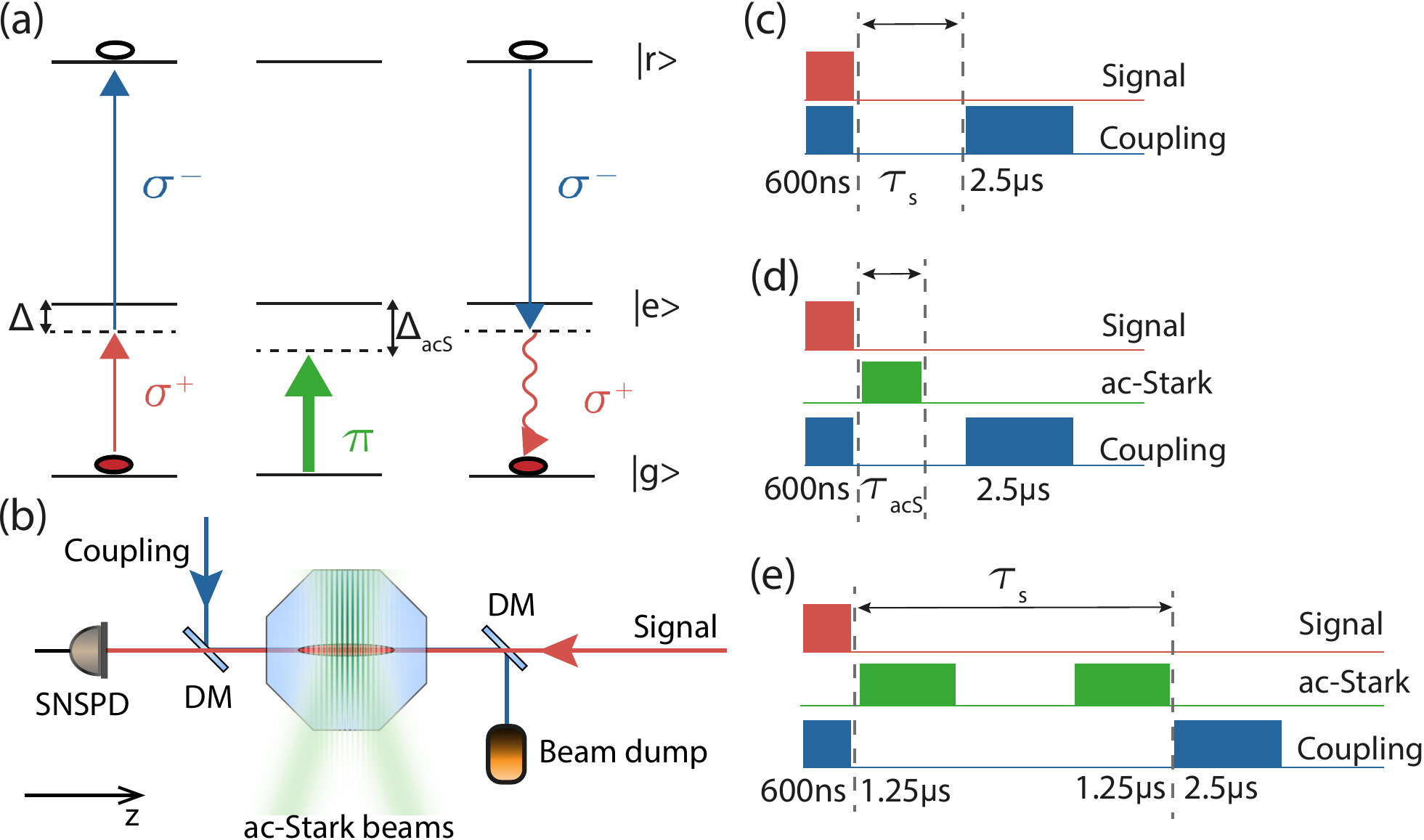}
\caption{\label{fig:schemat}\subfig{a} Relevant $^{87}\mathrm{Rb}$ energy level configuration. Atomic coherence is induced between $\ket{g}$ and $\ket{r}$ levels. 
\subfig{b} Experimental setup for presented protocol, based on ultracold atoms in a magneto-optical trap. The  signal laser passes excites a spin wave in the atomic cloud. The counter-propagating coupling laser is reflected from dichroic mirror (DM) and then introduced into the atomic chamber. During storage ac-Stark beams modulate the stored coherence. Later the spin wave is deexcited and the read-out is measured on superconducting nanowire single-photon detectors (SNSPD). 
\subfig{c} Experimental sequence for the measurement of the lifetime of unmodulated spin wave.
\subfig{d} Experimental sequence for the calibration of ac-Stark modulation impulses duration. 
\subfig{e} Experimental sequence for the measurement of the extended lifetime of modulated Rydberg spin wave.
} 
\end{figure}

The experiment is based on atomic memory that is built on rubidium-87 atoms trapped in a magneto-optical trap (MOT). The trapping and experiments are performed in a sequence lasting \SI{12}{\ms}, which is synchronized with mains frequency. Atoms form an elongated cloud in a cigar shape $(\SI{0.4}{\mm}\times\SI{0.4}{\mm}\times\SI{9}{\mm})$ with optical depth on the relevant atomic transition reaching 190, which would correspond to atomic number $N\approx10^8$ and atomic density $n\approx\SI{10e10}{\cm^{-3}}$. The ensemble temperature is \SI{78}{\micro\K}. After the cooling and trapping procedure, atoms are optically pumped to the state $\ket{g} \coloneqq 5^2S_{1/2}\, F = 2, m_F = 2$. We utilize the ladder system depicted in Fig.~\ref{fig:schemat}\subfig{a} to couple the light and atomic coherence. The experimental setup is presented in Fig.~\ref{fig:schemat}\subfig{b}. Signal laser with $\sigma^{+}$ polarization is red detuned by $\Delta = 2\pi\times\SI{40}{\mega\hertz}$ from the $\ket{g} \rightarrow \ket{e} \coloneqq 5^2P_{3/2}\,F = 3, m_F = 3$ transition. Counter-propagating coupling laser with $\sigma^{-}$ polarization is tuned to the resonance with signal beam enabling two-photon transition $\ket{g} \rightarrow \ket{r} \coloneqq 34^2D_{5/2}\,F = 4, m_F = 4$, inducing atomic coherence between $\ket{g}$ and $\ket{r}$ states. Blockade radius for $\ket{r}$ is $r_b=\SI{2.1}{\micro \m}$. We set waists of the coupling and signal beams in the cloud's near field to be respectively \SI{80}{\um} and \SI{70}{\um}. The ac-Stark modulation is performed with two $\pi$-polarized laser beams red detuned by $\Delta_\text{acS}$ = $2\pi\times\SI{1}{\giga\hertz}$ from the transition $\ket{g}\xrightarrow{} \ket{e}$. The beams are crossed at an angle $\theta = \ang{18.5}$ creating a sinusoidal interference pattern on the atomic medium. The beams are shaped by a set of cylindrical lenses to match the size of the atomic medium. The $\pi$ polarization of the beams reduces the maximal visibility of the interference to $\mathcal{V}\approx95\%$, which has a marginal effect when compared to other imperfections of the beams. We measure the read-out at the single-photon-level using superconducting nanowire single-photon detectors (SNSPD --- ID281).

\subsection{Sequence}
Our experiment consists of 3 parts, measurement of unmodulated Rydberg spin wave lifetime, calibration of modulating impulse duration and ultimately Rydberg spin wave lifetime extension protocol. Each of the parts can be divided into three stages:
(1) Writing the signal photons on atoms by inducing atomic coherence between levels $\ket{g}$ and $\ket{r}$ with signal and coupling impulses both lasting for $\SI{600}{\nano \s}$,
(2) Storing and modulating spin waves for time $\tau_s$,
(3) Read-out of stored signal with coupling impulse lasting for $\SI{2.5}{\micro \s}$ and subsequent detection with SNSPD.

To benchmark the efficiency of the lifetime extension protocol, we had to find the readout intensity of the unmodulated spin wave as a function of the storage time. In this part of the experiment, the modulating impulses are not illuminating the atoms. The sequence of this part of the experiment is presented on Fig. \ref{fig:schemat}\subfig{c}. By extending the storage time $\tau_s$ from 0 to $\SI{44}{\micro \s}$ we can acquire the readout intensity characteristics as a function of storage time. To the points that were measured this way, we fitted the Gaussian function in the explicit form $I(t) = I_0\exp{(-t^2/\tau_{\text{un}}^2 )}$ and obtained the value of unmodulated lifetime $\tau_{\text{un}} = \SI{2.4+-0.1}{\micro \s}$. 

Due to the geometry of our setup, zeroing out the Rydberg level spin-wave wavevector could be obtained only with the second order of the ac-Stark modulation. 
To perform the modulation with two interfering beams, we needed to find the optimal duration of the modulating impulses such that the acquired phase corresponds to the maximum of the $2^{nd}$ order Bessel function. To calibrate the optimal time for the impulses, we stored the light in the atoms according to the sequence presented in Fig. \ref{fig:schemat}\subfig{d}. In this protocol, during the second stage of the experiment, we illuminate the atoms with two beams that introduce spatially varying ac-Stark shifts. The time between the first and third stage of the sequence was set to $\tau_s = \SI{3}{\micro \s}$ while the duration of the modulating impulse $\tau_{acS}$ was changing from 0 to $\SI{3}{\micro \s}$. 
To the measured intensity of the retrieved signal as a function of the duration of modulating impulse, we fitted the square of the Bessel function of the $0^{th}$ order. With the parameters of the fitted function, we were able to determine that the optimal duration of the modulating impulses is $\tau_\text{acS} = \SI{1.25}{\micro\s}$, which corresponds to the maximal value of the $2^{nd}$ order Bessel function for our experimental parameters.

The final part of the experiment, which is a measurement of an extended lifetime, was performed according to the experimental sequence presented in Fig. \ref{fig:schemat}\subfig{e}.  
During the write-in stage the signal field, enabled by the coupling field, creates a short-lived spin-wave excitation between levels $\ket{g}$ and $\ket{r}$. The excitation is then modulated by two interfering ac-Stark beams for the optimal time $\tau_\text{acS} = \SI{1.25}{\us}$ resulting in second-order modulation of the spin wave. After modulation, the stored spin wave has a longer lifetime compared to the non-modulated spin wave. Finally, before the read-out stage, we repeat the modulation and retrieve the stored light with the use of the coupling laser.

\section{Results}
\begin{figure}[t]
\centering
\includegraphics[width = 1\columnwidth]{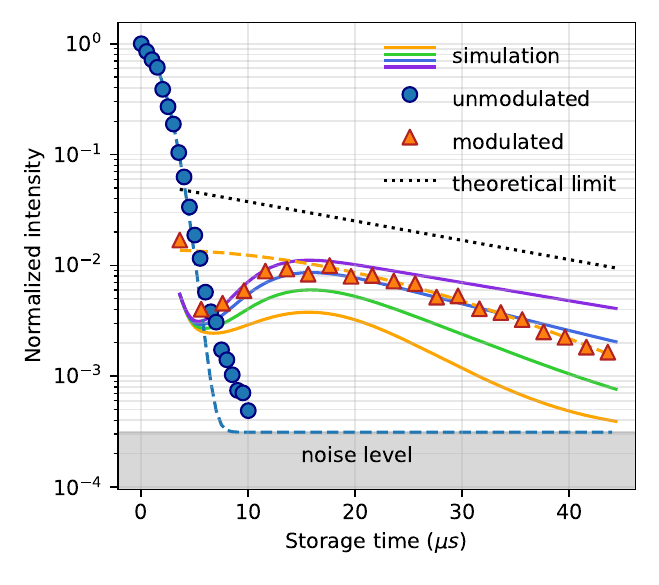}
\caption{\label{fig:lifetime} 
Normalized number of photons collected with SNSPD as a function of storage time. 
Blue dots represent the normalized intensity of unmodulated stored signal as a function of storage time. Orange triangles represent the normalized intensity of stored signal modulated by shortening its spin-wave wavevector. Dashed lines correspond to Gaussian function fitted to the experimental data with an added noise level. The dotted grey line is the maximal possible theoretical limit of the modulation efficiency calculated as $\eta(t) =  \eta_{\text{max}} \exp{(-\Gamma t)} $. 
The solid curves are the result of the numerical simulation described in the text with added noise level and
multiplied by $\eta_{\text{acS}}^2$ representing the imperfections of the ac-Stark beams. Each of the corresponds to different modulation wavevector $q$: violet - $q = 0.5k_0$, blue - $q = 0.485k_0$, green - $q = 0.471k_0$, orange - $q = 0.456k_0$.}

\end{figure}

The lifetime extension protocol was realized according to the sequence depicted in Fig.~\ref{fig:schemat}\subfig{e}. 
The photons for a single data point were collected for $\SI{155}{\s}$ in 7750 sequences of cooling and trapping atoms followed by write-in, storage of the signal as a Rydberg spin wave, and read-out with a repetition rate of $\SI{50}{\hertz}$. Every measured data point was normalized to the cumulative number of the detected photons for the storage time $\tau_s = 0$ in the case where no modulation was applied. 
The normalized number of photons of modulated and unmodulated stored signal is presented in Fig.~\ref{fig:lifetime}. 

The numerical simulation shows that the expected read-out efficiency after modulation will follow typical motional decay characteristics (Gaussian decay) after time $\approx 5 \tau = \SI{12}{\us}$, due to residual non-zero wavevector components of the atomic coherence. Following this observation, we fitted the Gaussian function to the data points on the right side of the dip in the intensity, that is for storage time $\tau_s \geq \SI{14}{\micro \s}$. The presence of the preliminary efficiency dip can be attributed to the mixing of residual components resulting from the modulation. We obtained the lifetime of the spin wave for the modulated signal $\tau_{\text{mod}} = \SI{27 +- 1}{\micro \s}$.
For the single realization of the experiment, the average number of photons for $\tau_{\text{s}} = 0$ of the unmodulated signal was $\bar{n} = \SI{5.7}{}$. 
The efficiency of single ac-Stark modulation is $\eta_{\text{acS}} = 71 \%$ which corresponds to the relative standard deviation of the intensity of the ac-Stark beams $\sigma = 19\%$ \cite{lipka_spatial_2019}. 

\section{Discussion}
In this article, we have demonstrated the method of prolonging the lifetime of the Rydberg level spin waves by an order of magnitude by introducing the ac-Stark lattice modulation in the quantum memory. 
The light shift is produced by two interfering beams far detuned from the atomic resonance. 
We have conducted the numerical simulation of the wavevector modulation to compare the experimental results with the theoretical predictions. 
The presented method is limited only by the radiative lifetime of a selected Rydberg state. 
The feasibility of the presented method allows for encoding long-lived qubits in atomic ensembles of cold atoms. 
We envisage that by combining our technique with state-selective modulation \cite{jiang_dynamical_2016}, the presented theoretical limit could be overcome, achieving storage efficiency close to the unity for a significant time.   
The current parameters of the experimental setup allow for the realization of the presented scheme, although they can still be improved. Increasing the the power of the ac-Stark beams would allow for reducing the time of the modulating pulses, leading to higher efficiency of the protocol allowing for saturation of the introduced theoretical limit for this modulation. 
Fully harnessing the potential of Rydberg lifetime extension protocol requires a cryogenic environment \cite{CantatMoltrecht2020,Muni2022} which would allow for decrease of the decoherence effects from black body radiation. 
The results of this article may open the route for many quantum information processing protocols. 
Moreover, the storage of long-lived Rydberg excitation introduces many prospects in the field of ultrasensitive electrometry, quantum computing and quantum simulation. 
\paragraph{Data availability}
Data for figure 3 has been deposited at \cite{Data2024} (Harvard Dataverse).

\begin{acknowledgments}
The “Quantum Optical Technologies” (MAB/2018/4) project is carried out within the International Research Agendas programme of the Foundation for Polish Science co-financed by the European Union under the European Regional Development Fund. This research was funded in whole or in part by National Science Centre, Poland grant no. 2021/43/D/ST2/03114. We thank K. Banaszek for generous support.
\end{acknowledgments}

\bibliographystyle{quantum}
\bibliography{refs}

\end{document}